# Partition function for two-dimensional nearest neighbour Ising model in the presence of external magnetic field

G. Nandhini and M.V.Sangaranarayanan[*]

Department of Chemistry

Indian Institute of Technology Madras

Chennai 600036 India

**Abstract**

The partition function for two-dimensional nearest neighbour Ising model in the presence of a magnetic field is derived. A comparison with the partition functions predicted by Onsager is carried out. The critical temperature estimated by two different methods yields good agreement with the result of Lee and Yang[1].

**Methodology**

Consider a two-dimensional nearest neighbour Ising model on a square lattice with the usual nearest neighbor Ising Hamiltonian ($H_T$) where J is the nearest neighbor interaction energy and N denotes the total number of sites, H being the external magnetic field.

The corresponding partition function is defined as[2]

$$Q = \text{Tr} \{\exp(-H_T/kT)\} \quad (1)$$

where k denotes the Boltzmann constant, T being the absolute temperature. We have derived Q employing a novel method of defining the system parameters as

$$Q_{(H \neq 0)} = \left[ \frac{(e^{(2J+H/2)/kT} + e^{(4J+H)/kT}) + \sqrt{(e^{(2J+H/2)/kT} + e^{(4J+H)/kT})^2 - 4(e^{(6J+3H/2)/kT} - e^{(9J/2+3H/2)/kT})}}{2} \right]^{N/2}$$

(2)

[*] sangara@iitm.ac.in

For H = 0, the above partition function reduces to

$$Q_{(H=0)} = \left[ \frac{(e^{2J/kT} + e^{4J/kT}) + \sqrt{(e^{2J/kT} + e^{4J/kT})^2 - 4(e^{6J/kT} - e^{9J/2kT})}}{2} \right]^{N/2} \quad (3)$$

The prediction of the above equation with that due to Onsager[3] is shown in Fig 1 for various values of J/ kT. A very good agreement can be noticed.

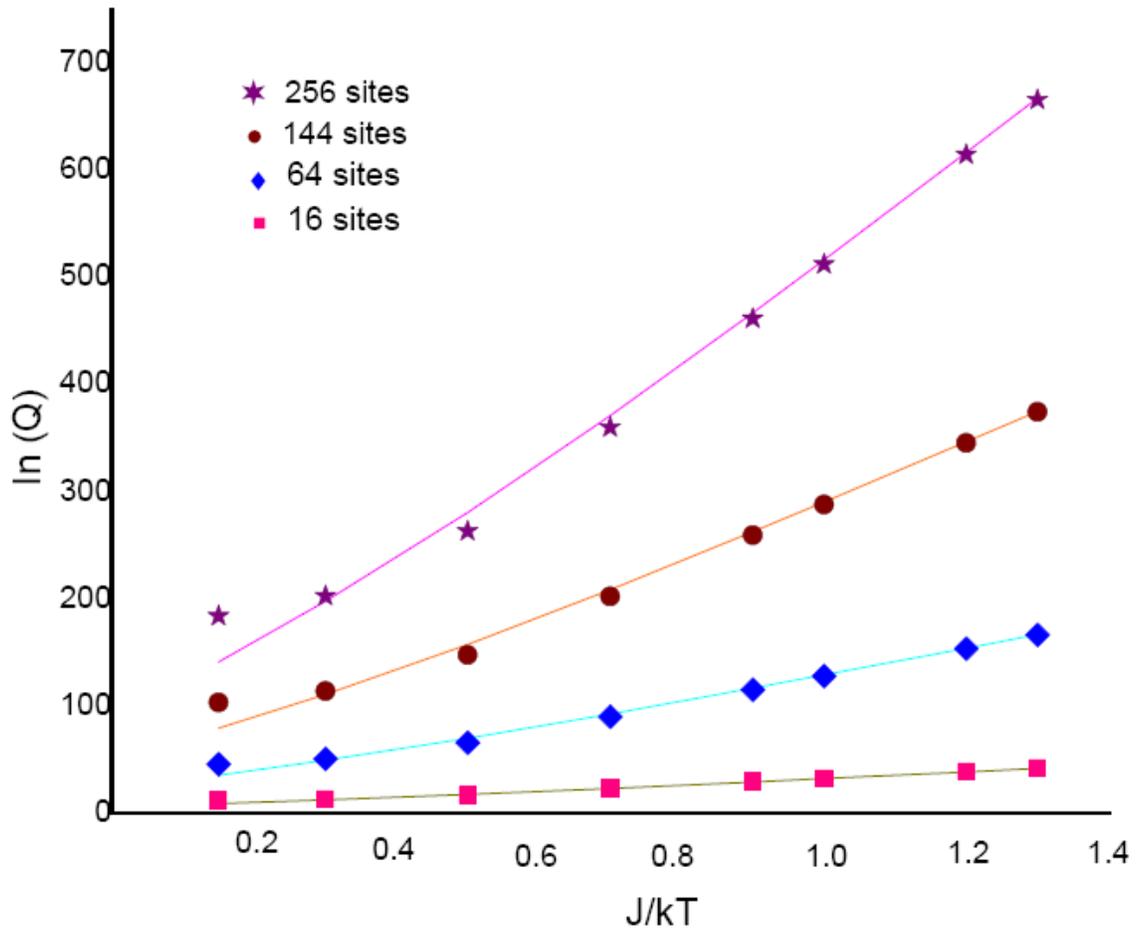

*Fig 1 - A comparison of the partition function values estimated from our equation with that of Onsager's analysis. Line denotes the estimates obtained by us, while the symbols denote the values obtained from Onsager's exact solution.*

Since no explicit result for the partition function valid for all J and H is hitherto available, a precise analysis to verify the validity of our eqn is precluded at present.

**Spontaneous Magnetization**

The precise definition of the spontaneous magnetization for two-dimensional Ising model is subtle[4]. A commonly employed prescription is

$$M_{H=0} = \lim_{N \to \infty} \frac{1}{N} \left( \frac{d \ln(Q)}{dH} \right)_{H \to 0^+} \quad (4)$$

where the order of taking limits is crucial. While a complete analysis of magnetization will be discussed elsewhere, we derive $\frac{1}{N}\left(\frac{d\ln(Q)}{dH}\right)$ as

$$\frac{1}{N}\left(\frac{d \ln(Q)}{dH}\right) \propto \frac{(e^{(4J+H)/kT} + \frac{e^{(2J+H/2)/kT}}{2}) + \frac{(e^{2(4J+H)/kT} + \frac{e^{(4J+H)/kT}}{2} - \frac{3e^{(6J+3H/2)/kT}}{2} + 3e^{(9J/2+3H/2)/kT})}{\sqrt{(e^{(4J+H)/kT} + e^{(2J+H/2)/kT})^2 - 4(e^{(6J+3H/2)/kT} - e^{(9J/2+3H/2)/kT})}}}{[(e^{(4J+H)/kT} + e^{(2J+H/2)/kT}) + \sqrt{(e^{(4J+H)/kT} + e^{(2J+H/2)/kT})^2 - 4(e^{(6J+3H/2)/kT} - e^{(9J/2+3H/2)/kT})}]}$$

(5)

and,

$$\frac{1}{N}\left(\frac{d \ln(Q)}{dH}\right)_{H=0} \propto \frac{(e^{4J/kT} + \frac{e^{2J/kT}}{2}) + \frac{(e^{2(4J)/kT} + \frac{e^{4J/kT}}{2} - \frac{3e^{6J/kT}}{2} + 3e^{(9J/2)/kT})}{\sqrt{(e^{4J/kT} + e^{2J/kT})^2 - 4(e^{6J/kT} - e^{9J/2/kT})}}}{[(e^{4J/kT} + e^{2J/kT}) + \sqrt{(e^{4J/kT} + e^{2J/kT})^2 - 4(e^{6J/kT} - e^{(9J/2)/kT})}]}$$

(6)

$J/kT_c$ is deduced as 0.4409 which is in excellent agreement with the value of 0.4407 arising from the analysis of Lee and Yang[1].

During the search for critical temperatures arising from the magnetization expression, an entirely novel method of deducing the same from the partition function itself was found by us which yields $J/kT_c = 0.4438$. This value is in excellent agreement with the value of 0.4407 expected for the critical temperature when the spontaneous magnetization becomes zero.

In Summary, the partition function for the two-dimensional nearest neighbour Ising model in the case of a non-vanishing magnetic field is derived. When the magnetic field is zero, the partition functions predicted by Onsager's exact solution are obtained. A new method of estimating the critical temperature is proposed.